\documentclass{article}

\usepackage{PRIMEarxiv}

\usepackage[utf8]{inputenc} % allow utf-8 input
\usepackage[T1]{fontenc}    % use 8-bit T1 fonts
\usepackage{hyperref}       % hyperlinks
\usepackage{url}            % simple URL typesetting
\usepackage{booktabs}       % professional-quality tables
\usepackage{amsfonts}       % blackboard math symbols
\usepackage{nicefrac}       % compact symbols for 1/2, etc.
\usepackage{microtype}      % microtypography
\usepackage{lipsum}
\usepackage{fancyhdr}       % header
\usepackage{graphicx}       % graphics
\usepackage{amsmath}
\usepackage{subfig}

\graphicspath{{media/}}     % organize your images and other figures under media/ folder

\title{Genetic Features for Drug Responses in Cancer - Investigating an Ensemble-Feature-Selection Approach
}

\author{
  Johannes Schlüter \\
  Universität Bielefeld \\
  Bielefeld\\
  \texttt{j.schlueter@uni-bielefeld.de} \\
   \And
   Alexander Schönhuth \\
  Universität Bielefeld  \\
  Bielefeld\\
  \texttt{aschoen@cebitec.uni-bielefeld.de} \\
}

\begin{document}
\maketitle

\begin{abstract}
%% Text of abstract
Predicting drug responses using genetic and transcriptomic features is crucial for enhancing personalized medicine. In this study, we implemented an ensemble of machine learning algorithms to analyze the correlation between genetic and transcriptomic features of cancer cell lines and IC50 values, a reliable metric for drug efficacy. Our analysis involved a reduction of the feature set from an original pool of 38,977 features, demonstrating a strong linear relationship between genetic features and drug responses across various algorithms, including \textit{SVR}, \textit{Linear Regression}, and \textit{Ridge Regression}. Notably, copy number variations (CNVs) emerged as more predictive than mutations, suggesting a significant reevaluation of biomarkers for drug response prediction. Through rigorous statistical methods, we identified a highly reduced set of 421 critical features. This set offers a novel perspective that contrasts with traditional cancer driver genes, underscoring the potential for these biomarkers in designing targeted therapies. Furthermore, our findings advocate for IC50 values as a predictable measurement of drug responses and underscore the need for more data that can represent the dimensionality of genomic data in drug response prediction. Future work will aim to expand the dataset and refine feature selection to enhance the generalizability of the predictive model in clinical settings.
\end{abstract}
% keywords can be removed

%%Graphical abstract

\centering  
\includegraphics[width=\textwidth,height=0.75\textheight,keepaspectratio]{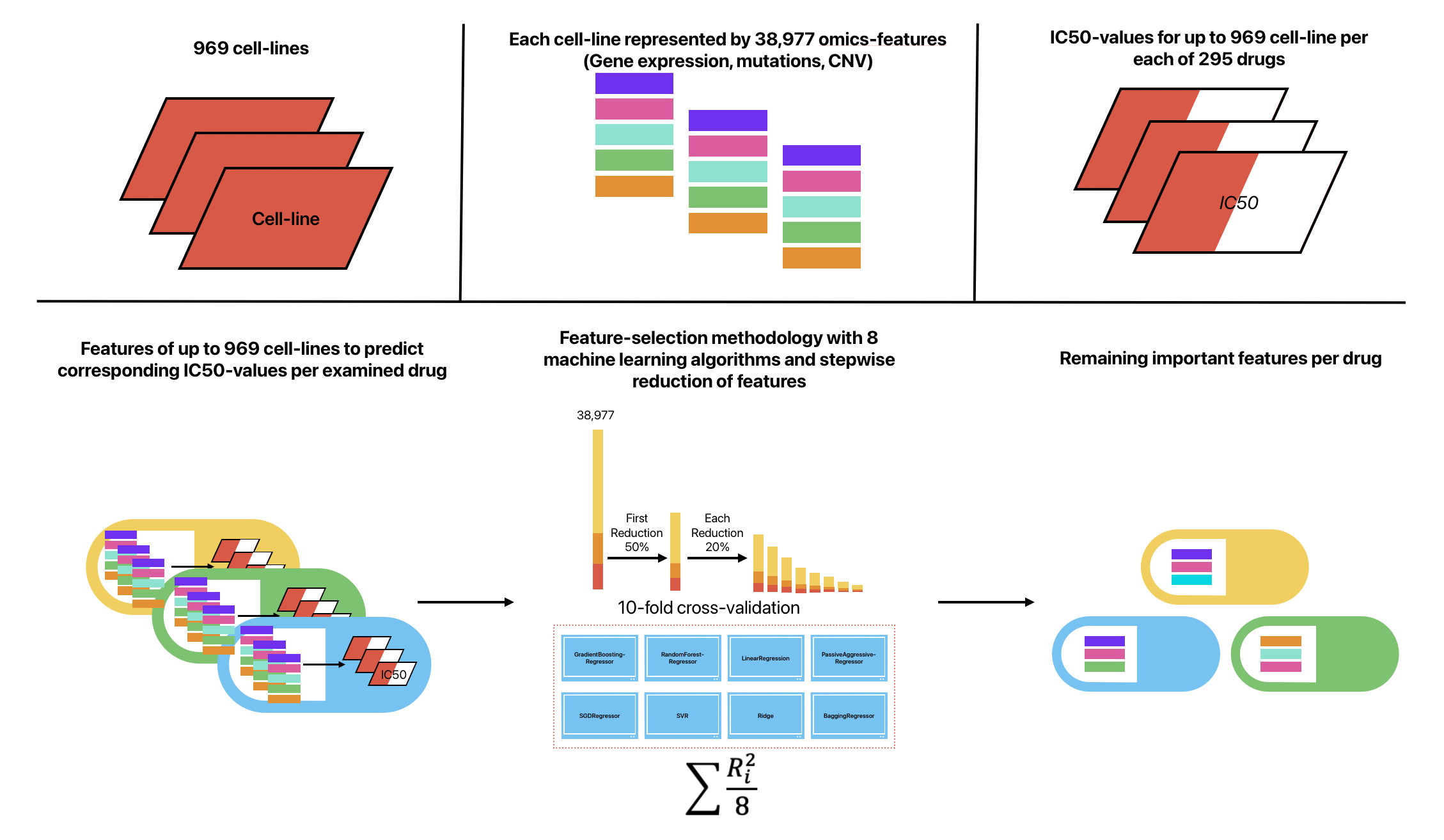}
\label{fig_graphical_abstract}
%\includegraphics{grabs}

%% Add \usepackage{lineno} before \begin{document} and uncomment 
%% following line to enable line numbers
%% \linenumbers

%% main text
%%

%% Use \section commands to start a section
\section{Introduction}
\label{sec1}
%% Labels are used to cross-reference an item using \ref command.
Cancer is the second most frequent cause of death globally and is object of permanent research~\cite{Sung2021}.
Despite significant advancements, our understanding of this disease or, better, set of related diseases has not been completed yet, but new approaches under utilization of new technologies such as Artificial Intelligence (AI) or modern pharmacotherapy yield towards an increasingly personalized and individualized medicine~\cite{Kourou2015}. Promising as this may be, a deep understanding of underlying mechanisms is missing. The sheer number of genes and their varied expressions within a single cell evoke computational challenges, not to mention the additional complexity introduced by mutations and other complex changes~\cite{ComputationalComplexitySCStegle2015,ComplexityOmicsCortés-Ciriano2022,ComplexityMutationsCNVStratton2009}. Nonetheless, to address these challenges, we have adapted an ensemble-feature-selection approach initially proposed by Lopez-Rincon et al.~\cite{LopezRincon2020}. Our aim is to investigate the learnability of drug sensitivity based on genetic features and to identify key gene features. By leveraging this approach, we hope to facilitate research in this challenging field and contribute to the broader understanding of cancer treatment.

Personalized medicine tailors treatment to the individual characteristics of each patient, often considering genetic, environmental, and lifestyle factors~\cite{PersonlizedMedizineMathur2017}. In the context of cancer, this approach is particularly promising because it allows for treatments that are more effective and less harmful than traditional one-size-fits-all therapies~\cite{SINGH2024100860}. However, achieving this level of customization requires a deep understanding of the genetic basis of cancer and how it influences drug response~\cite{Chin2011}.

One of the main challenges in cancer treatment is predicting how different patients will respond to specific drugs. This variability in drug response is largely due to genetic differences between individuals~\cite{DEPALMA2012111}. Identifying the genetic features that influence drug sensitivity is crucial for developing personalized treatment plans~\cite{cancers15153837}. For instance, certain mutations can increase or decrease cancer cells' susceptibility to specific therapies.~\cite{GeneFeaturesPersonalizedTreatment}.

The complexity of cancer genetics necessitates advanced computational methods to analyze the vast amount of data generated by genomic studies~\cite{Libbrecht2015}. Machine learning, a subset of AI, has emerged as a powerful tool in this regard. It can handle large datasets and identify patterns that may not be apparent through traditional statistical methods~\cite{BigDataMachineLearningYoo2014-st, MLInsteadOfStatistic7906512}. By using machine learning algorithms, researchers can predict drug responses and identify potential biomarkers for cancer therapy~\cite{BiomarkerIdentificationInCancerMLJagga01082015, FeatureSelectionDrugResponseAli2019}.

Ensemble methods in machine learning combine multiple models to improve prediction accuracy and robustness~\cite{Dietterich2000}. In our study, we adapted an ensemble-feature-selection approach from Lopez-Rincon et al.~\cite{LopezRincon2020} to analyze genetic data from cancer patients. This method allows us to identify the most relevant genetic features that influence drug sensitivity, improving our understanding of the genetic basis of drug response in cancer.

One well-studied dataset in the field of drug responses in cancer is the Genomics of Drug Sensitivity in Cancer (GDSC) dataset, which provides extensive data on the drug responses of cancer cell lines and their genomic information \cite{Yang2013}. This dataset has been widely used for applying predictive models. Additionally, the Cell Model Passport (CMP) offers comprehensive genetic, transcriptomic, and proteomic information about the investigated cancer cell lines, aiding in the understanding of underlying mechanisms~\cite{CMPvanWeerden2020}. Studies utilizing the GDSC dataset, such as those by Iorio et al. (2016), have highlighted various genetic markers associated with drug response \cite{Iorio2016}. 
Similarly, the Cancer Cell Line Encyclopedia (CCLE) by Garnett et al. (2012) links specific genetic alterations to drug sensitivities, providing valuable insights for personalized medicine \cite{Garnett2012CellModelPassport}.
Additionally, methods such as CRISPR-Cas9 screens have been employed to identify essential genes that modulate drug sensitivity, further refining our understanding of gene-drug interactions \cite{Hart2015}.

Recent studies have continued to build on this foundational work. For instance, Wang et al. (2022) used machine learning models to predict drug response based on multi-omics data, demonstrating the potential of integrated approaches for personalized medicine in cancer~\cite{Wang2022}. Moreover, Malik et al. (2021) explored the use of deep learning models to predict drug sensitivity from genomic profiles, highlighting the advancements in computational approaches~\cite{Malik2021}.
Building on these developments, recent transformer-based and graph convolutional neural network (GCN) approaches, such as swNet, DeepTTA, and DeepAEG, have achieved high accuracy in predicting drug responses on the dataset investigated in this study~\cite{DEEPTTA10.1093/bib/bbac100, SWNETZuo2021, DEEPAEGLao2024}. However, while these models excel in predictive performance, they provide limited insights into the underlying biological and genetic mechanisms driving drug response.
%Tang et al. (2021) developed a novel framework combining multi-omics data and machine learning to improve the accuracy of drug sensitivity predictions~\cite{Tang2021}. Similarly, Zhao et al. (2022) proposed an integrative approach using graph neural networks to leverage the complex relationships within multi-omics data for better drug response predictions~\cite{Zhao2022}.

%Considering the dataset which has been investigated by this work, latest approaches based by on transformers and graph convolutional neural network such as swNet, DeepTTA and DeepAEG predicted drug responses for this data considerably accurately~\cite{DEEPTTA10.1093/bib/bbac100, SWNETZuo2021, DEEPAEGLao2024}. However, they do not offer any insights in the underlying biological and genetic concepts. 

Exploring the roles of different gene types, such as oncogenes and tumor suppressor genes, versus genes responsible for drug sensitivity, is vital for understanding drug resistance mechanisms. Research indicates that while oncogenes and tumor suppressor genes are crucial for cancer development and progression, the genes influencing drug sensitivity might be distinct. Studies such as those by Vogelstein et al. (2013) have delved into the genetic underpinnings of cancer, highlighting the differences in gene function and their implications for treatment~\cite{Vogelstein2013}. More recent work by Bailey et al. (2018) has expanded on this, identifying numerous driver mutations and their roles in various cancers~\cite{Bailey2018}. In addition, recent studies by Dakal et al. (2024) have provided insights into the differential roles of oncogenes and tumor suppressor genes, emphasizing the complexity of genetics in cancer~\cite{Dakal2024}.

Unlike these studies, which often focus on individual gene-drug interactions, our approach integrates feature selection and ensemble learning to identify a comprehensive set of predictive features across multiple drugs, potentially uncovering broader patterns of drug sensitivity. Our method employs an iterative reduction process combined with ensemble models to refine the feature set, ensuring that the selected features are not only statistically significant but also robust across different drugs. This holistic approach contrasts with the more gene-centric analyses typically conducted in previous studies, providing a more integrated view of the genetic determinants of drug sensitivity.

Recent studies have also explored drug response prediction using matrix factorization and drug-centric modeling. For example, Emdadi and Eslahchi~\cite{TCLMF} proposed a logistic matrix factorization method (TCLMF), which models drug sensitivity as a binary outcome based on latent representations learned from cell line data. While gene expression is used indirectly to map tumor samples to similar cell lines, it is not used as direct input for prediction or feature selection. Similarly, Yassaee and Eslahchi~\cite{MinDrug} developed MinDrug, a method that predicts IC50 values for unseen drugs based on a small representative subset of reference drugs, without incorporating any molecular features.

In contrast to these approaches, our method is designed specifically for interpretable feature selection. By applying an ensemble learning strategy to gene expression, mutation, and CNV data, we aim to identify robust and biologically meaningful genomic features associated with drug response, rather than focusing on predictive performance alone. This enables a clearer understanding of which features may mechanistically influence sensitivity and resistance.

In this manuscript, we explain (i) how we obtained and preprocessed the data, (ii) how we adapted the approach from Lopez-Rincon et al. to our needs and (iii) that our results demonstrate valuable findings, highlighting significant genetic features that influence drug sensitivity. These features should be considered for further research and differ from common strategies and investigations. 

As described, numerous other studies have explored the GDSC2 database using various methods. However, our approach seeks to address several crucial questions that had been remained unanswered in prior work:

\begin{enumerate}
    \item \textit{Which genes determine drug sensitivity?} Identifying specific genes that influence how cancer cells respond to different drugs is crucial for developing targeted therapies~\cite{GenesForTargetedTherapyLEE2018188}. \label{Quest:WhichGenes}
    \item \textit{Can we predict the outcome of drug administration based on a small subset of genetic features?} Predictive models can help clinicians choose the most effective treatments for individual patients, potentially improving outcomes and reducing side effects~\cite{PredicteModelForClinicalDecisionMaking}. \label{Quest:PredictionFromSubset}
    \item \textit{Do genetic features vary depending on the drug?} 
    Understanding whether different drugs are influenced by different or the same genetic features can inform the development of combination therapies and personalized treatment plans~\cite{GenesForTargetedTherapyLEE2018188}.\label{Quest:DiversityAcrossDrugs}
    \item \textit{Do oncogenes and tumor suppressor genes differ from the genes responsible for drug sensitivity?} Exploring the roles of different types of genes can provide insights into the mechanisms of drug resistance and sensitivity~\cite{RolesOfGenesWang2019-ps}.\label{Quest:ComparisonOncoTumorsupressor}
\end{enumerate}

Through our research, we aim to provide clarity on these questions, ultimately contributing to the development of more effective and personalized comprehensible cancer treatments.

%Section text. See Subsection \ref{subsec1}.

%% Use \subsection commands to start a subsection.
%\subsection{Example Subsection}
%\label{subsec1}

%Subsection text.

%% Use \subsubsection, \paragraph, \subparagraph commands to 
%% start 3rd, 4th and 5th level sections.
%% Refer following link for more details.
%% https://en.wikibooks.org/wiki/LaTeX/Document_Structure#Sectioning_commands

%% Refer following link for more details.
%% https://en.wikibooks.org/wiki/LaTeX/Mathematics
%% https://en.wikibooks.org/wiki/LaTeX/Advanced_Mathematics

\section{MATERIALS AND METHODS}

\subsection{Data and Preprocessing}
\begin{table}[ht]
\centering

\caption{Summary of data characteristics in this study.}
\label{tab:dataset_summary}
\resizebox{\textwidth}{!}{%
\begin{tabular}{||l|l||}
\hline
\textbf{Property} & \textbf{Value / Description} \\
\hline
Dataset Source & GDSC2 (drug response), CMP (genomic features) \\
Number of drugs & 295 \\
Drug response metric & IC50 \\
IC50 distribution & Range: [-8.769011, 13.847363], mean: 2.83, std: 2.76 \\
Number of cell lines per drug & 969 (varies per drug: min: 225, max: 969, median: 892, mean: 819.46) \\
Gene expression features & 37,602 (encoded as TPM) \\
Mutation features & 612 (one-hot encoded by gene) \\
Copy Number Variation (CNV) features & 763 (integer count per gene) \\
Total number of features & 38,977 \\

\hline
Processing note & Separate model trained per drug due to disjoint cell lines \\
\hline
\end{tabular}%
}
\end{table}

We obtained cell lines and IC50 values to the corresponding 295 drugs from the GDSC2 dataset from the Sanger institute~\cite{Yang2013}. The genetic information to cell lines was taken from the CMP~\cite{CMPvanWeerden2020}. The CMP provides 37,602 gene expression features, 612 different mutations and 763 different CNV (38,977 features in total) for each of the 969 cell lines. Meta-information of the used data is summarized in Table~\ref{tab:dataset_summary}
The number of cell lines $C_d$  tested differs for each drug $d \in D$. There is no set $C$ such that the intersection of all sets of cell lines $C_i$ is not the empty set or in other words: The intersection of all sets of cell lines from all drugs $d \in D$ is the empty set. 

\[|\bigcap_{d \in D} C_d|=\{\}\]

Therefore, we performed our method for each drug separately.

%#\begin{align*}
%#A\cup B&=\{x |\;x\in A\lor x\in B\}\\
%#A\cap B&=\{x |\;x\in A\land x\in B\}\\[.5em] \limsup_{n\to\infty} a_n
%#&=\sup \{x\in\R|\; \forall n\in\N\;
%#\exists m\in\N:m\ge n\land x<a_m\}\\
%#&=\sup \bigcap_{n=1}^\infty \bigcup_{m=n}^\infty (-\infty,a_m)
%# \end{align*}

The gene expression values we utilized were encoded as transcripts per million (TPM): 
\[TPM_i = \frac{q_i/l_i}{\sum_j(q_j/l_j)}*10^6\] 
\noindent
where \(q_i\) is the reads which are mapped to a transcript and \(l_i\) is the length of the transcript. This encoding seeks to make samples comparable~\cite{Zhao2021}. We one-hot-encoded the 612 mutations such that we stored the name of the gene in which the mutation possibly occurs. If the dataset provides information about a mutation,
%where $M$ is the set of all in the dataset recorded mutations for the corresponding cell line we set the value  $m_i = 1$, if there was no mutation found for that cell line we set $m_i = 0$:
\[B_ij = \begin{cases}
1, & \text{if mutation in feature in column j for cell line in row i in B} \\
0, & \text{otherwise}

\end{cases}\] 

This encoding is based on the assumption that a gene either influences a drug’s response or remains inactive, depending on its functionality. However, a mutation within a gene may change the genes functionality. Thus, a mutation may either change the drug response or not. Since we investigate the drug response determining genes, a one-hot-encoding seems elegant and promising. 
\\
For the 763 CNVs, we utilized the number of CNVs within a gene as a single integer value as provided by CMP. Integrating more detailed information may be useful in future research, but for an investigation of the importances of genes, this approach reduces the complexity of the feature space while maintaining focus on gene relevance to pharmacodynamics and pharmacokinetics.

The sets are separated by drugs. Thus, for each drug, there are between 225 and 969 IC50 values. This depends on the number of cell lines on which the drug was tested. \\
For each cell line, there are 38,977 features. Therefore, the matrix we learn can be of maximal size 969 (cell lines) times 38,977 (features) and the corresponding label vector of size 1 times 969 dependant on the number of cell lines that have been tested for each specific drug. It is important to note that the ratio of features to samples (cell lines) is not favorable, as the feature space is high-dimensional while the number of samples is approximately 40 times smaller. 

%To do so, we applied a changed version of the ensemble-feature-selection-method developed by Lopez et al.~\cite{LopezRincon2020} adapted to our needs.

%Then we created two datasets: One for a possibly good performing feature selection, where we did not divide the set into training- and test-set to keep as many of the few samples as possible. 

%The other one was divided into a training-set (80\% of all samples) and a test-set (20\% of all samples). This was done for each drug separately since there is no intersection of samples for all drugs. We created a test-set to investigate whether our model is not only able to extract relevant features, but if we can also create a recommendation system already by this. 

%To do so, we applied a changed version of the ensemble-feature-selection-method developed by Lopez et al.~\cite{TODO} adapted to our needs.

\subsection{Algorithm}

\begin{figure}[ht!]

\centering
    \subfloat[\centering Iterative Feature Reduction]{{\includegraphics[width=5cm]{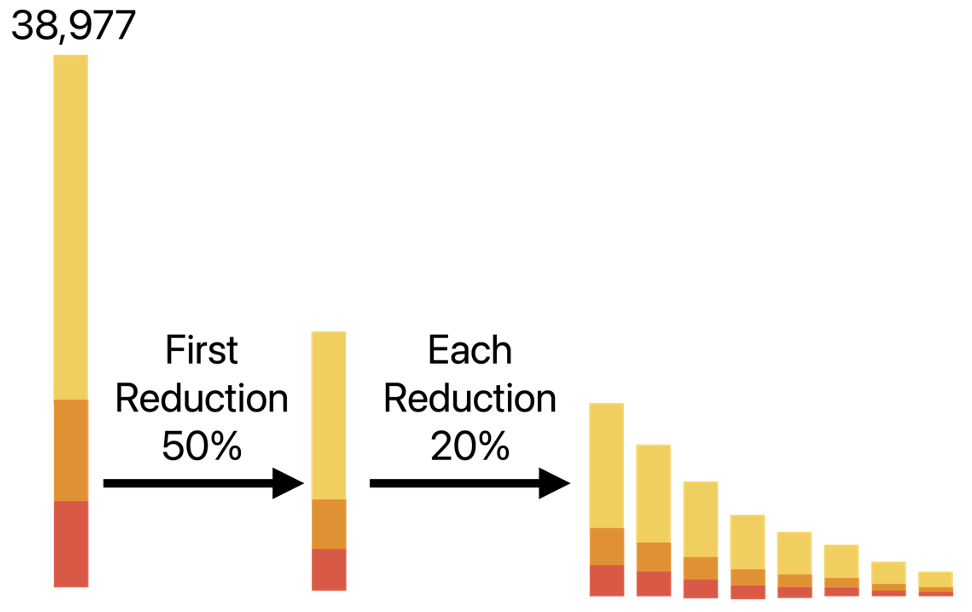} }}%
    \qquad
    \subfloat[\centering Adapted Ensemble Approach]{{\includegraphics[width=5cm]{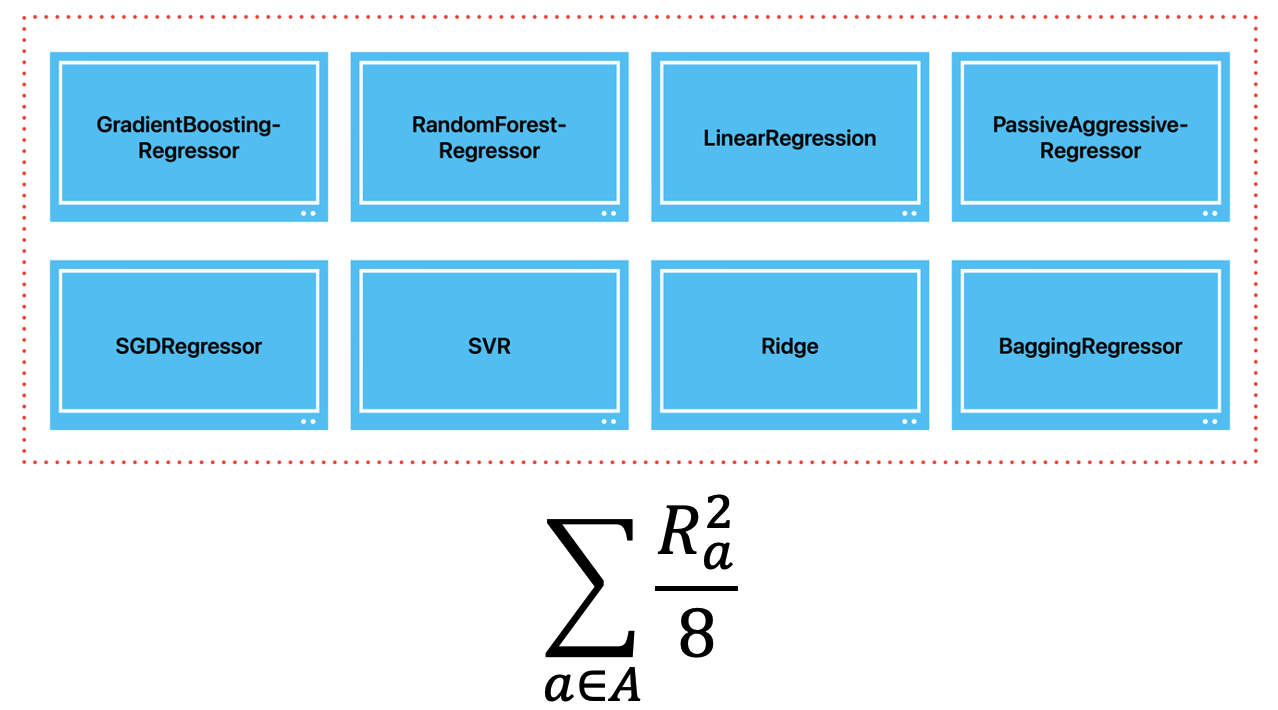} }}%
    \caption{a) Scheme of the iterative reduction procedure (colors indicating the different types of features: yellow: gene expression (the largest amount of features by far); orange: CNVs; red: mutations;). Beginning with an initial 50\% cut, we reduce the features by 20\% for each iteration thereafter until the feature set is empty. b) The adapted ensemble approach from Lopez et al.~\cite{LopezRincon2020}. Let $A$ be the set of algorithms in our ensemble approach. We evaluated the performance of the approach by dividing the sum of the resulting $R_a^2$ by the number of algorithms $\lvert A \lvert=8$.}%
    \label{fig:EnsembleApproach}%
%\centering  
%\includegraphics[width=0.5\columnwidth]{images/Feature_elimination.png}
%\caption{a) Scheme of the iterative reduction procedure (colors indicating the different types of features: yellow: gene expression (the largest amount of features by far); orange: CNVs; red: mutations;). Beginning with an initial 50\% cut, we reduce the features by 20\% for each iteration thereafter until the feature set is empty. b) The adapted ensemble approach from Lopez et al.~\cite{LopezRincon2020}. Let $A$ be the set of algorithms in our ensemble approach. We evaluated the performance of the approach by dividing the sum of the resulting $R_a^2$ by the number of algorithms $\lvert A \lvert=8$.  }
%\label{fig_reduction}

%\begin{center}
%\includegraphics[width=0.5\columnwidth]{images/FeatureSelection.png}
%\end{center}
%\caption{
%The adapted ensemble approach from Lopez et al.~\cite{LopezRincon2020}. Let $A$ be the set of algorithms in our ensemble approach. We evaluated the performance of the approach by dividing the sum of the resulting $R_a^2$ by the number of algorithms $\lvert A \lvert=8$. 
%}
%\label{fig:EnsembleApproach}
\end{figure}
As outlined in Figure~\ref{fig:EnsembleApproach}, our analysis utilized an ensemble feature selection method, a backward-feature-elimination, adapted from Lopez et al.~\cite{LopezRincon2020}, tailored to handle the regression nature of our problem. This approach has been shown to be a powerful tool for feature selection, especially, in the field of biomedicine and biomarker detection. Furthermore, it is able to work with limited data where deep learning approaches tend to fail or overfit while it preserves the ability to model linear and non-linear relationships in the data. Finally yet importantly, it enables us to determine interpretable features.
\sloppy We deployed a suite of eight machine learning algorithms from scikit-learn, including \textit{GradientBoostingRegressor}, \textit{RandomForestRegressor}, \textit{LinearRegression}, \textit{PassiveAggressiveRegressor}, \textit{SGDRegressor}, \textit{SVR}, \textit{Ridge}, and \textit{BaggingRegressor} (Figure~\ref{fig:EnsembleApproach} (b)). 
These algorithms were chosen based on the work from Lopez et al.~\cite{LopezRincon2020} whereas this is a regression problem instead of a classification problem and we, therefore, replace logistic regression by linear regression as it is a classifying algorithm although the name may suggest differently. 
The other algorithms are exchanged with their corresponding Regressor. The full code is available on \url{https://github.com/JHelge/Genetic-Features-for-Drug-Responses-in-Cancer}.

The regression problem lies in predicting the IC50 score for each drug separately. Let $D$ be the set of all drugs of size $n=295$ from GDSC. Thus, we actually have very similar 295 regression problems. For each $d\in D$ we have a set of cell lines $C_d$ of size $m_d$ for which the drug was tested. $m_d$ varies across all drugs $d$ in a range of 223 and 969. 
This is represented as a label-vector $v_d$ of length $m_d$ for each drug for each regression problem where each entry represents an IC50-score for the corresponding cell line for this specific drug. Again, for each cell line $C_d$ there is a set of genetic and transcriptomic features $G_c$ of size $o=38,977$. Then, for each drug $d\in D$ we create the feature matrix $B_d$ of size $m_d*o$ containing the values for the corresponding features from $G_c$ as columns and the cell lines from $C_d$ as rows.
As our approach is an ensemble feature selection approach, we try to obtain an optimal feature set from the original feature set (38.977) by backwards elimination.
Each algorithms starts with the whole feature set (38.977) and trains against the drug's corresponding IC50. We apply a 10-fold cross-validation to validate the models' performances. During the training, we try to optimize for the \(R^2(y,\hat{y}) = 1 - \frac{\sum\limits_{c\in C_d}(y_c-\hat{y_c})^2}{\sum\limits_{c\in C_d}(y_c-\bar{y_c})^2} \) as the default metric for these algorithms. Each algorithm provides direct or indirect score to access the feature importances to the obtained result. 
We take the feature importances and calculate the averaged feature importances across all algorithms for each feature. Then, we sort all features by their averaged importance and reduce the feature space by 50\% initially and 20\% per subsequent iteration by removing the least important features (Figure~\ref{fig:EnsembleApproach} (a)).
The aggressive initial cut eliminating 50\% of features was arbitrarily chosen to simply decrease the necessary computational resources. 
In future work, the effect of different reduction step sizes could be further investigated to optimize this approach. However, the original approach from Lopez et al. started with an even more aggressive cut reducing the number of feature to 1000 at once~\cite{LopezRincon2020}. 
Taking into account the large initial number of features in our dataset, we decided not to reduce the number of features to 1000 directly. Subsequent reductions of features were executed until the set of remaining feature set is an empty set. 
Thus, for 38,977 features we executed 40 reductions.

We save the subsets from all reduction-steps. Then, we average the performances of all algorithms to obtain the best performing subset. This, undergoes another cross-validation to examine our model's stability. The best set and evaluation-criteria are saved as well.

%This is done with both sets, the training-set without a test-set and the other training-set for which we hold back 20\% of the data as a test-set. 

To evaluate the features we obtained not only by performance of our model, but through context sensitive biological and medical analysis and interpretation, we calculated the overlap or Szymkiewicz–Simpson coefficient of the best performing subsets between all drugs:

\[overlap(S_{d_i},S_{d_j})=\frac{|S_{d_i}|\cap |S_{d_j}|}{\min(S_d{_i}|,|S_{d_j}|}\]

Where $S_{d_i}$ and $S_{d_j}$ are best performing subsets of the corresponding feature sets $G_{d_i}$ and ${G_{d_j}}$ for two distinct drugs $d_i$ and $d_j$ respectively. Moreover, we applied a consensus scoring to rank the selected features.

%Then, we also calculated the overlap within a gene, thus, for a best-performing subset whether genes exist that where found as an important feature in terms of expression and CNV, of expression and mutation, of mutation and CNV or of all three of them. 

%To test the predictive performance of our approach on new data and to build a recommendation system, we utilized the feature sets, we obtained from the training-set where we created a corresponding test set. We used the best-performing set of features to train our model and, then, test it on the untouched test-set. We did this for each drug separately, since there is no intersection of cell lines for all drugs as explained above. 

%We evaluate our approach utilizing the \textit{R\textsuperscript{2}} as a common metric in machine learning with good comparability. 

By taking a detailed look at our data, we noticed that even though each drug-ID is only assigned once, there are eight drugs, which appear twice with different IDs but the same name-label. It is mentioned that drugs with IDs below 1.000 were screened by at the Massachusetts General Hospital (Boston, USA) and IDs greater than 1.000 were screened by the Wellcome Sanger Institute (Cambridge, UK). However, the GDSC2 assigns IDs greater than 1.000 only. These Drugs are shown in Table~\ref{Tab:TableDoppelganger}. Since these duplicates have the same name, we assume them to have similar IC50 values for the same cell lines and we will compare the feature sets created by our model to show the power of this approach as we expect our tool to obtain similar feature sets, but also withdraws on the basis of the biological complexity.

\section{RESULTS}

%\begin{figure*}[thb]
%\centering  
%\includegraphics[width=\textwidth]{images/Averaged Algso Reductions.png}
%\caption{The progression of the averaged \textit{R\textsuperscript{2}} across drugs for each algorithm over iterations of reduction.}
%\label{AvAlgPer}
%\end{figure*}
\subsection{Algorithms and Performances}
\begin{figure}
\centering  
\includegraphics[width=0.8\columnwidth]{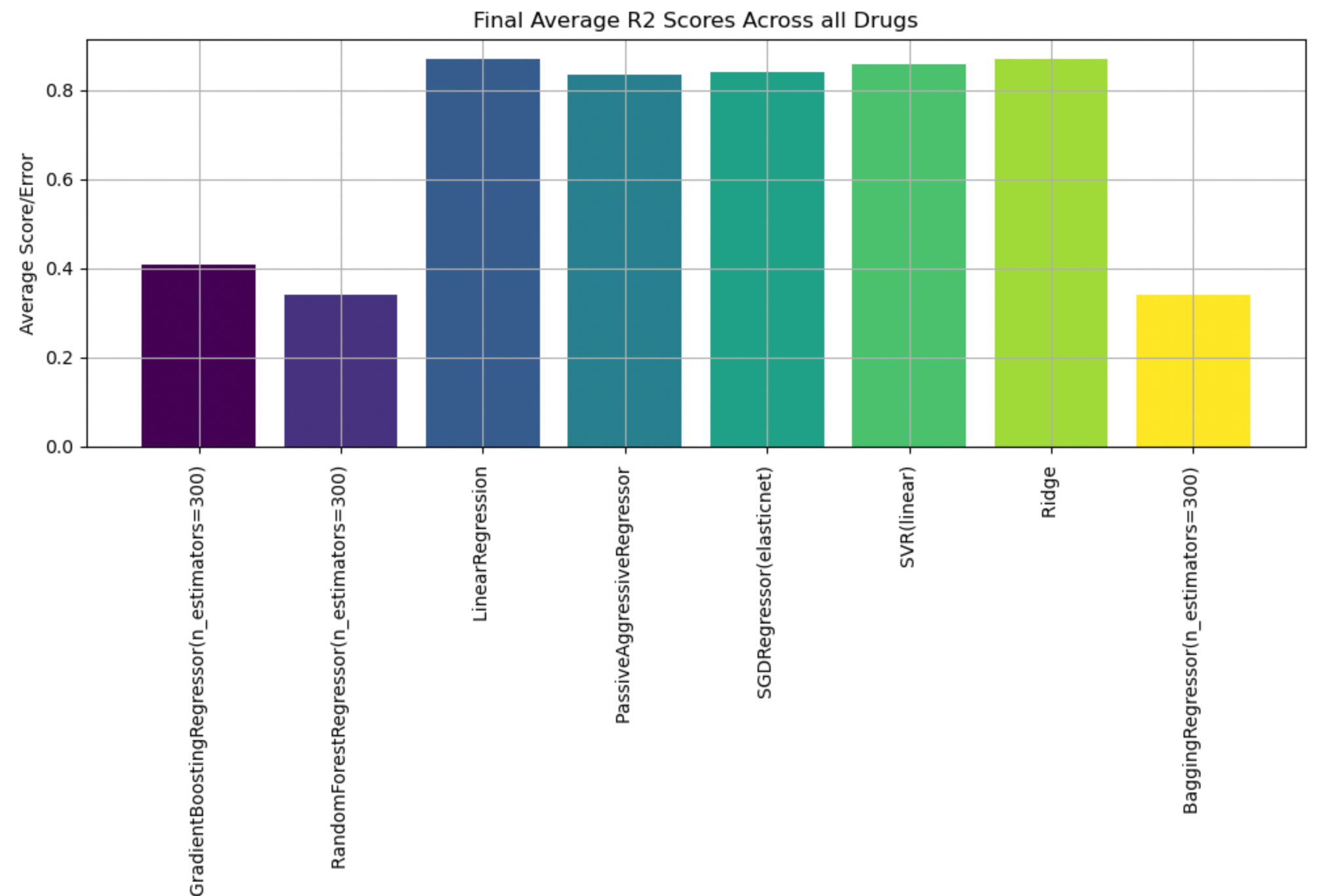}
\caption{The average performance across all drugs for each algorithm on the best performing feature set where each feature set was determined for a particular drug.}
\label{fig_sim}
\end{figure}

We measured the performance of the single algorithms of our approach by the \textit{R\textsuperscript{2}} and as shown in Figure~\ref{fig_sim} we obtained average \textit{R\textsuperscript{2}}s between 0.342 (\textit{BaggingRegressor(n\_estimators=300)}) and 0.872 (\textit{Ridge}). The average performances of all algorithms averaged over all drugs was 0.671. In comparison, the \textit{R\textsuperscript{2}} without reducing the number of features is $-1$ for any drug while the \textit{R\textsuperscript{2}} with feature elimination is 0.671 on average across all drugs and algorithms. To mitigate the impact of extreme negative outliers in \textit{R\textsuperscript{2}} scores, we clipped all \textit{R\textsuperscript{2}} values below $-1$ to $-1$. 
This conservative normalization allows for a fairer comparison of feature selection strategies without overemphasizing model failures. Without this clipping, the average \textit{R\textsuperscript{2}} across all drugs was approximately $-1.46*10^{27}$.
In addition to comparing against models using all features, we also evaluated a simpler feature selection baseline based on Pearson correlation. For each drug $d$, we selected the top-$\lvert S_d \lvert$ features with the highest absolute correlation to the drug response, where $\lvert S_d \lvert$ equals the number of features selected by our ensemble method for that drug. Using the same regressors and 10-fold cross-validation, this correlation-based approach yielded an average \textit{R\textsuperscript{2}} of $-0.0419$, compared to $0.6689$ achieved by our ensemble selection strategy. In none of the 295 drugs did the correlation-based selection outperform the ensemble method. This result emphasizes the advantage of our strategy in identifying robust, multivariate and biologically meaningful predictors of drug sensitivity.

%Figure~\ref{AvAlgPer} shows the averaged course of the algorithms' performances throughout the 40 reduction steps. We notice the very similar course of \textit{SVR}, \textit{LinearRegression} and \textit{Ridge} and that these three are among the best performing algorithms together with \textit{SGDRegressor}. 

\begin{figure}[ht!]
\centering  
\includegraphics[width=0.8\columnwidth]{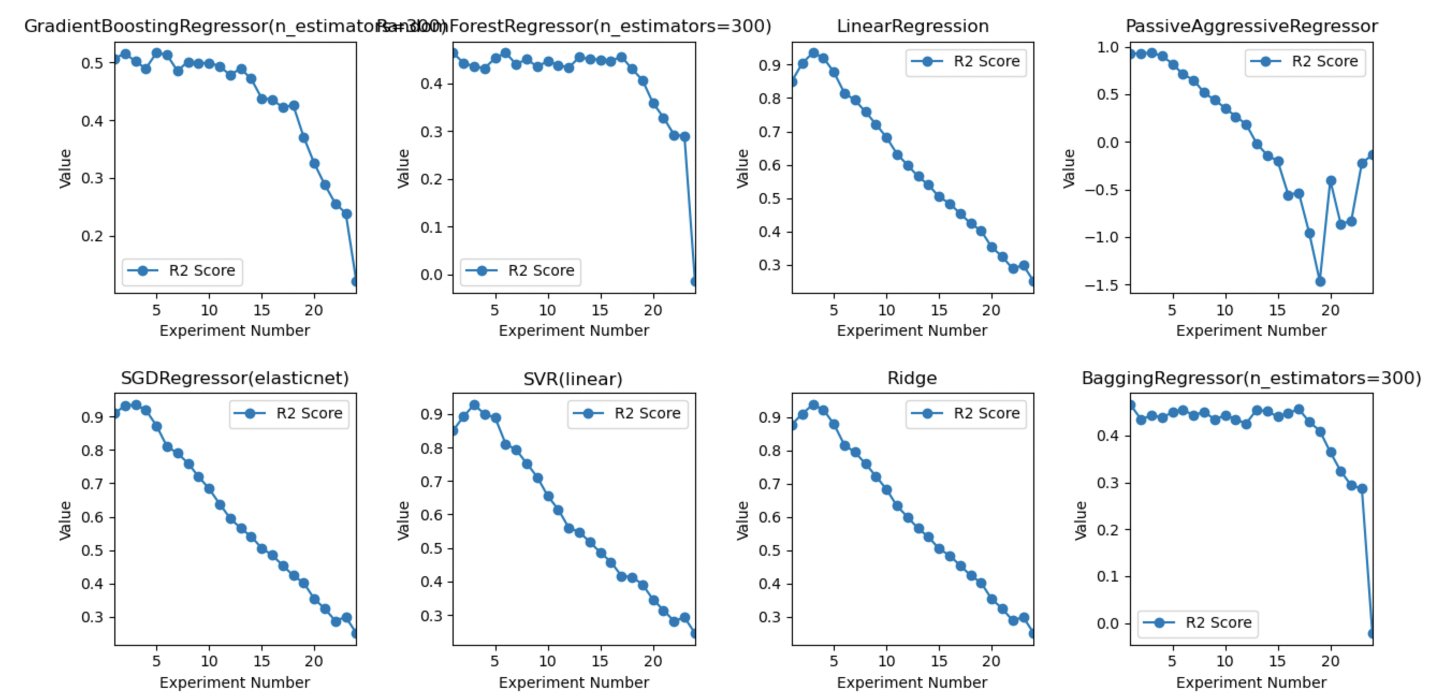}
\caption{An exemplary development of the performance of each algorithm across the last 30 iterations for Acetalax (ID: 1803).}
\label{singleCourseAlgo}
\end{figure}

In Figure~\ref{singleCourseAlgo} we observe the descent of the single algorithms performances after reaching an optimum. \textit{SGDRegressor}, \textit{SVR}, \textit{Ridge} and \textit{Linear Regression} have an almost linear descent from the optimum, whereas, other algorithms descent differently and perform less well except \textit{PassiveAggressiveRegressor}.

\subsection{Features}

Depending on the drug $d$, we obtained a best performing feature set $S_d$ according to the averaged \textit{R\textsuperscript{2}} from our ensemble feature selection. Let $n=295$ be the total number of drugs.

\begin{figure}[h]
\centering  
\includegraphics[width=0.6\columnwidth]{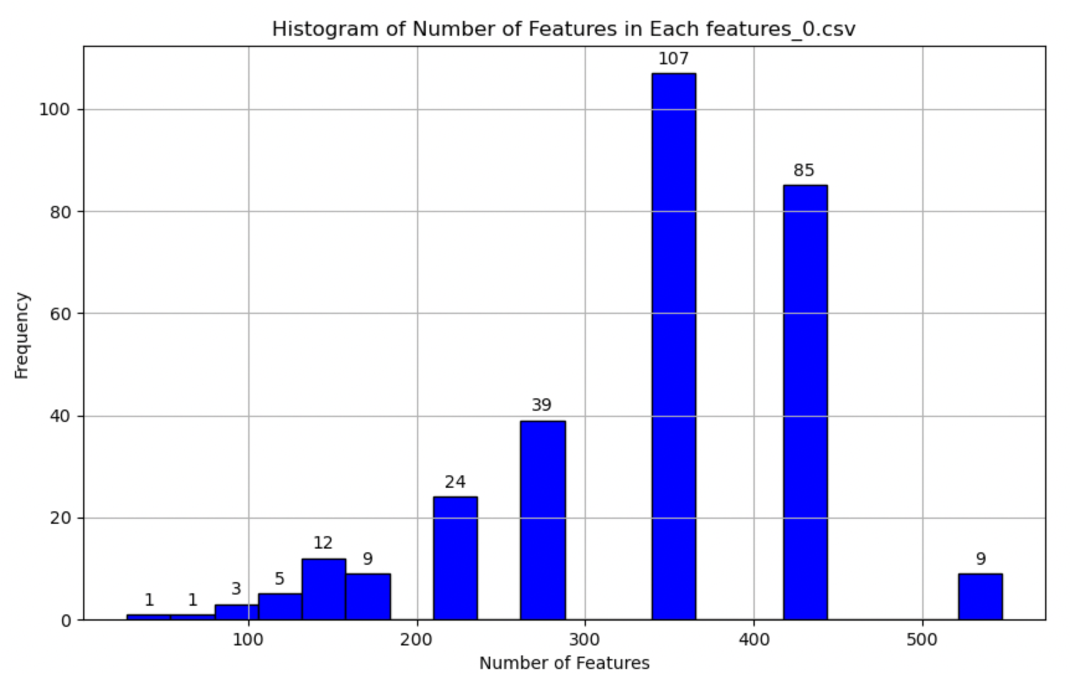}
\caption{The distribution of best performing feature sets. The smallest set is of size 28 and the largest of size 557.}
\label{FeatureSetSizes}
\end{figure}

We obtain a set size $\lvert S_d \lvert$ between 28 and 557 features (see Figure~\ref{FeatureSetSizes}) for the best performing feature sets of each drug. Thus, it may not seem surprising that the intersection of all $n$ obtained feature sets $S_d$ is the empty set:

\[|\bigcap_{d \in D} S_d|=\{\}\]

Still, we were hoping to find a possibly small union of all feature sets. However, the size of the union of all $n$ feature sets $S_d$ is equal to 20,429 which is more than 50\% of the total number of features (38,977):

\[|\bigcup_{d \in D} S_d|=20,429\]

\begin{figure}
\centering  
\includegraphics[width=0.6\columnwidth]{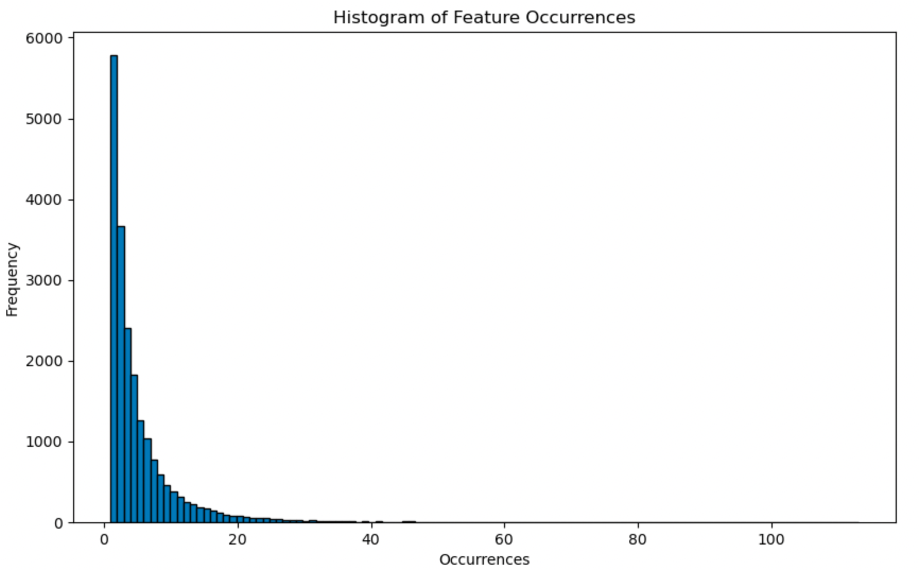}
\caption{Histogram of the occurrences of all features from the best performing sets.}
\label{FeaturesAll}
\end{figure}

\begin{figure}
\centering  
\includegraphics[width=0.9\columnwidth]{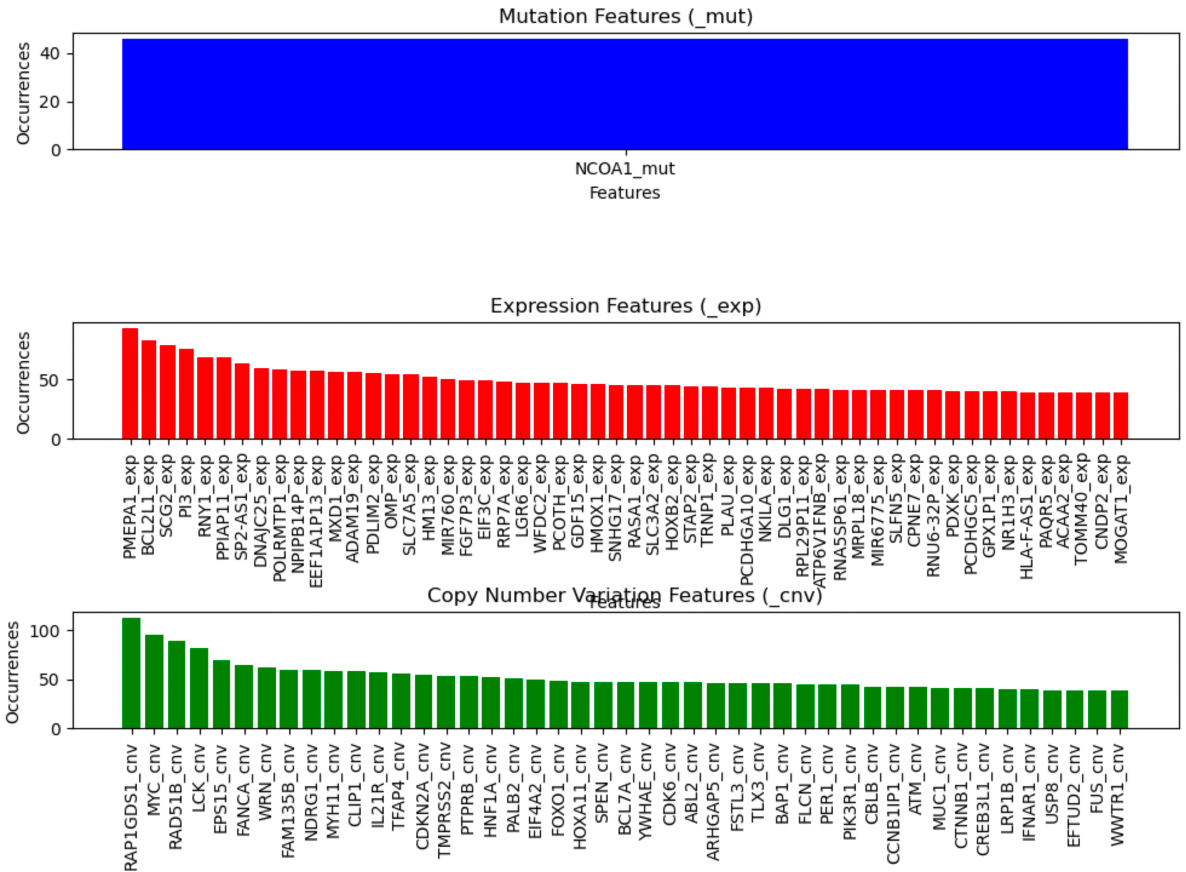}
\caption{The 100 most frequent features across the best performing sets separated by feature type.}
\label{Features100}

\includegraphics[width=0.9\columnwidth]{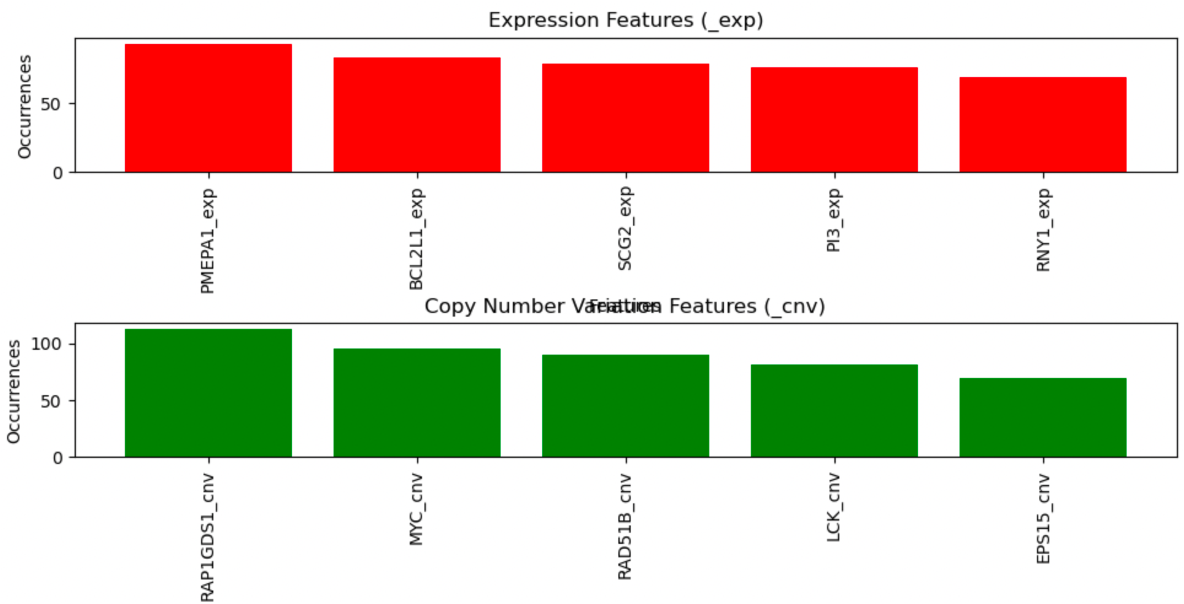}
\caption{The 10 most frequent features across the best performing sets separated by feature type.}
\label{Features10}
\end{figure}

By looking at Figure~\ref{FeaturesAll} we can observe that most of the features occur only once or twice if we look at all feature sets $S_d$ for all $n$ drugs. Figure~\ref{Features100} and \ref{Features10} show the features which were among the 100 most frequent features and 10 most frequent features respectively, separated by type of feature. Within the 100 most frequent features, 54 genes are found because of their expression, 45 because of their CNV, but only one mutation was found. Among the 10 most frequent features, we find 5 CNV features and 5 gene expression features.

\subsection{Comparison with known cancer drivers}
To access the importance of known cancer drivers (oncogenes and tumor-suppressors) to drug responses, we compared our most frequent features to two lists of driver genes from CMP and COSMIC~\cite{CMPvanWeerden2020, COSMIC10.1093/nar/gky1015}.
Comparing the 10 most frequent features with the driver genes from CMP and the consensus from COSMIC we see that those 5 genes and only those genes which are included as CNVs are found in both resources. 
Taking a look at the 100 most frequent features, we rediscover 41 features from COSMIC and 47 features from CMP. 

\subsection{Feature Occurrences in Computed Feature Sets}
Since we observed almost half of all selected features (9,442) to occur at most twice we calculated the probability of features to occur at least in as many feature sets as we found them. We keep those features which have a significant impact according to the following test procedure and removed all features, below this threshold. 
We compute the p-values and apply a Bonferroni correction as a conservative correction of the significance level. We choose this conservative approach to include important features only, regardless that less frequent features may be very important for certain drugs as well. The Bonferroni correction is applied by dividing the significance level ($p=0.05$) by the number of experiments. This is done to take into account that a high number of experiments increases the probability of observing a result by chance. In this case, the high number of comparisons increases the probability to consider features important just by chance. We decrease the significance level from $p=0.05$ to $p=\frac{0.05}{k}$ where $k$ is the number of experiments. Since we test each possible feature $g \in G$, where $G$ is the set containing all features we perform $o=38,977$ experiments. Thus, the corrected significance level is $p=\frac{0.05}{38,977}=1.283*10\textsuperscript{-6}$. \\

Then, we calculated the corresponding p-values for each feature. For this, we took the number of occurrence for each feature across all feature sets. 
As the number of features differs between all feature sets, we assume that our feature sets were all of size 557 which is the largest feature set we obtained. Thus, the p-value would be even larger for smaller sets. Otherwise, we would need to compute for each feature separately the p-value depending on the size of the feature sets in which it is included. This represents a very careful selection of features to avoid features without greater meaning, but with the risk of excluding few possibly relevant features.
We found that from 24 occurrences on, the p-value is below the corrected significance level. Then, we removed all features that occurred in less than 24 sets. 

Finally, we ended up, having 421 features in our possibly relevant feature set. This feature set includes 144 driver genes from CMP and 115 from COSMIC.

\begin{table}[ht]
\centering

\caption{Significant enrichment of biological processes and pathways.}
\label{tab:enrichment_results}

\resizebox{\textwidth}{!}{%

\begin{tabular}{||l|l|l|r||}
\hline

\textbf{Source} & \textbf{ID} & \textbf{Name} & \textbf{p-value} \\

\hline

KEGG & KEGG:05166 & Human T-cell leukemia virus 1 infection & 0.0079  \\
GO:BP & GO:2001233 & Regulation of apoptotic signaling pathway & 0.0137  \\
KEGG & KEGG:05220 & Chronic myeloid leukemia & 0.0382  \\
\hline

\end{tabular}

}
\end{table}

To evaluate the biological interpretability of the selected features, we performed pathway enrichment analysis using g:Profiler on the top 10 most frequently selected genes across all drugs. 9 out of 10 features could be mapped to KEGG or GO pathways (90\% annotation rate), indicating strong biological characterization.
Three cancer-relevant pathways were significantly enriched ($p~<~0.05$), namely \textit{Human T-cell leukemia virus 1 infection} (KEGG:05166, p~=~0.0079, 60\% precision), \textit{regulation of apoptotic signaling pathway} (GO:2001233, $p~=~0.0137$, 44\% precision), and \textit{Chronic myeloid leukemia} (KEGG:05220, $p~=~0.0381$, 40\% precision) as presented in Table~\ref{tab:enrichment_results}. These results demonstrate that even a minimal feature set derived from our model aligns with well-established cancer-related biological processes, lending interpretability and credibility to the selection strategy.

\subsection{Duplicates}

For the eight duplicate-pairs, which exist with two different IDs in the data, we obtain different feature-sets. To put the overlap of the duplicate-pairs into context we compute the probability that two sets of the corresponding sizes show an overlap of at least the obtained size. 
Let $P(X)$ be the probability of the event $X$. Let $o$ be the total number of features we have at the beginning. Let $r$ and $s$ be the number of features of the feature sets we obtained for the two drugs we compare and let $t$ be the number of overlapping elements. It holds:
\[P=\frac{\binom{t}{o}\binom{n-t}{o-t}\binom{r-o}{o-s}}{\binom{s}{o}\binom{r}{o}}\]

The probabilities can be found in Table~\ref{Tab:TableDoppelganger}.

We also compute the corresponding test-statistic for the hypergeometric test and apply a conservative Bonferroni correction. The significance level is, according to this, decreased from $p=0.05$ to $p=\frac{0.05}{k}$ where $k$ is the number of experiments. Since we compare each drug $d_i \in D$, where $D$ is the set containing all Drugs, with all remaining drugs we perform $n=\frac{\vert D\vert \textsuperscript{2}+\vert D\vert}{2}=43,660$ experiments.
Thus, the corrected significance level is $p=\frac{0.05}{43,660}=1.145*10\textsuperscript{-6}$. Finally, we perform  the hypergeometric test for the most probable overlap of the duplicate-pair from Table~\ref{Tab:TableDoppelganger} and return a p-value of $p=9.507*10\textsuperscript{-7}$, thus, being smaller than the corrected significance level.
Therefore, it is significantly unlikely to obtain the overlap of feature sets by chance. However, the feature sets still differ, but differences may represent the complexity of the biological and pharmacological context.

\begin{table}[ht]
\caption{Table of all drugs which where tested twice in the dataset and the corresponding intersection of the two best performing feature sets for this drug.}
\centering

\begin{tabular}{||c c c c||}

\hline
Name & ID One & ID Two  & Intersection\cr
\hline
Acetalax  & 1803 & 1804 & 118 \cr 

Bleomycin & 1378 & 1812 & 32 \cr

Dactinomycin & 1811 & 1911 & 21\cr

Docetaxel & 1007 & 1819  & 21\cr

Fulvestrant & 1200 & 1816 & 17\cr

Oxaliplatin &  1089 & 1806 & 39\cr

Selumetinib & 1062 & 1736 & 17\cr

Uprosertib & 1553 & 2106 & 32\cr

GSK343 & 1553 & 2106 & 12\cr

Ulixertinib & 1553 & 2106 & 26\cr

\hline

    \end{tabular}
    \label{Tab:TableDoppelganger}

\end{table}
\section{DISCUSSION}

%\subsection{Key Findings}
In this study, we have investigated an ensemble feature selection approach to identify relevant genetic features for predicting drug responses in cancer. While recent research has focused on deep learning models like transformers for drug response prediction, the identification of key genetic predictors has received comparatively less attention. Moreover, as multiomics data have become increasingly relevant in computational genomics, we explore the significance of diverse omics data types, including mutations, CNVs, and gene expression data, for drug response prediction tasks.

Our approach achieves reasonably strong R\textsuperscript{2} values despite not being optimized for predictive performance through hyperparameter tuning. In future work, our approach may benefit from task-specific hyperparameter tuning, despite the average performance already being reasonably strong without.
We successfully reduce the original feature set (38,977 features) by two to three orders of magnitude, resulting in feature sets ranging between 28 (1.434\%) and 557 (0.072\%) features. This substantial reduction in feature space demonstrates the efficiency of our method. Interestingly, feature sets show a significant overlap for drugs that were tested twice under different conditions in the dataset, reinforcing the robustness of our approach. Among the identified features, CNVs emerged as particularly crucial, while the number of significant mutational features remained relatively small. After applying Bonferroni correction, we identified a set of 421 genetic features that are statistically significant for drug response prediction. Notably, these features differ in part from known cancer driver genes.

%\subsection{Comparison with Existing Methods}
Contrary to recent studies that utilize modern non-linear methods, our findings suggest that the relationship between genetic features and drug responses is predominantly linear. While transformer-based models offer enhanced predictive capabilities, our results indicate that linearity plays a major role in most cases. As provided in our work, the linear algorithms seem to push the performance of our feature selection towards smaller feature sets. Also, the average performance of the linear algorithms is greater than the performance of non-linear algorithms. This leads us to the assumption that genetic features influence drug responses primarily linearly rather than non-linearly. This has practical implications for personalized medicine, where explainability is essential for legal reasoning, patient education, and therapeutic decision-making. Deep learning approaches, such as DeepTTA and swNet, have demonstrated the power of non-linear models in genomics. However, our study focuses on feature selection rather than maximizing predictive performance. The identified features suggest that only a small subset of genetic markers is sufficient for drug response prediction, reducing computational requirements and improving model interpretability in cancer therapy.

Furthermore, our results highlight the importance of CNVs in predicting drug responses, suggesting that the predictive power of mutational data may be overestimated. This aligns with consensus findings from CMP and COSMIC, which emphasize the relevance of CNVs in cancer research. Given that COSMIC primarily focuses on mutations in cancer, our gene-oriented approach provides a broader perspective by incorporating multiple genomic alterations (mutations, CNVs, and gene expression changes) that can influence gene function and drug response. In comparison to gene expression data, the high influence of CNVs may be due to the fact that CNVs eventually result in greater expression, but may result in a larger and usually crucial change in a gene's expression while gene expression is noisier and consists of mostly irrelevant features. In regard to mutations, CNVs are more representative for the actual outcome of transcription and translation while mutations do not always refer to defect products, and may still fulfill their role partially.

%\subsection{Limitations and Future Directions}
While our findings offer valuable insights, our study is limited by the relatively small number of cell lines per drug. The dataset does not exceed 969 cell lines, and in many cases, the sample size is even smaller. Additionally, our gene-oriented approach may not fully capture all biological effects, as it depends on data acquisition and preprocessing methods that could introduce biases or errors. Such limitations are common in real-world genomic datasets. Although our study identifies significant genetic features, further validation with larger datasets is necessary to assess the generalizability of our selected feature sets.

Future research should focus on experimental validation of the identified feature sets through in vitro and in vivo studies. These genetic features should be of particular interest in follow-up experiments that aim to elucidate underlying biological mechanisms. Understanding these mechanisms will further clarify the relationship between genetic alterations and drug responses. Additionally, drug development may benefit from prioritizing drug response-associated genes over traditional cancer driver genes, as the genetic determinants of tumor growth and survival are not necessarily the same as those influencing drug response in cancer cells.

%\subsection{Conclusion}
In relation to our initial research questions, we identified a set of 421 significantly relevant features for drug response prediction (Question~\ref{Quest:WhichGenes}). Our findings suggest that a subset of genetic features is sufficient for predictive tasks and that the relationship between genetic alterations and drug responses is largely linear (Question~\ref{Quest:PredictionFromSubset}). However, the relevant feature subset varies across different drugs (Question~\ref{Quest:DiversityAcrossDrugs}), shifting attention away from known oncogenes and tumor-suppressor genes while highlighting the greater predictive value of CNVs (Question~\ref{Quest:ComparisonOncoTumorsupressor}). This underscores the need for a more comprehensive approach to genomic data analysis in drug response prediction, where CNVs should not be overlooked in favor of mutational data alone.

\section{CONCLUSION}

This study provides several key insights into the relationship between drug responses and genetic as well as transcriptomic features.

First, our findings confirm IC50 values as a reliable metric for drug response prediction, supported by observed linear correlations with genetic features. This reinforces the idea that, despite the increasing use of complex non-linear models, linear relationships play a predominant role in many cases.

Second, our results suggest that CNVs play a more crucial role in drug response mechanisms than previously assumed, while mutational data, though essential for cancer development, appear to be of lesser importance for drug response prediction. This finding challenges the common focus on mutations in predictive modeling and underscores the need for a broader genomic perspective.

Third, we find that a strong emphasis on cancer driver genes is only partially useful for predicting drug responses. Instead, our analysis highlights a set of significantly relevant features that may be more directly associated with drug efficacy, shifting attention from tumor development mechanisms to treatment response.

For future research, we aim to develop a practical recommendation system that leverages the identified feature sets for more effective drug response predictions. Additionally, expanding the dataset and incorporating more diverse biological contexts will be crucial to improving the generalizability of our findings. Further experimental validation is necessary to clearly distinguish between cancer driver genes and genes specifically relevant for drug response, thereby refining our understanding of genomic determinants in precision medicine.

\section*{Declaration of generative AI and AI-assisted technologies in the writing process}
During the preparation of this work the author(s) used ChatGPT in order to assist in refining the clarity and correctness of the text, particularly with respect to grammar, vocabulary usage, and stylistic consistency. The tool was utilized as a supplementary aid to enhance the overall readability and coherence of the manuscript. After using this tool/service, the author(s) reviewed and edited the content as needed and take(s) full responsibility for the content of the publication.

%Bibliography
\bibliographystyle{unsrt}  
\bibliography{templateArxiv}

\end{document}